\documentclass[aps,nofootinbib,twocolumn,superscriptaddress,showpacs,,floatfix,10pt]{revtex4}
\usepackage{amssymb,amsbsy}
\usepackage{amsmath,graphicx,color,epsfig}
\usepackage{slashed}
\usepackage{relsize}
\usepackage{axodraw}

\definecolor{purple}{rgb}{0.5,0,0.5}
\definecolor{blue}{rgb}{0.0,0,0.9}
\usepackage[colorlinks=true, pdfstartview=FitV, linkcolor=purple, citecolor= purple, urlcolor=blue]{hyperref}

\begin{document}

\title{Pseudoscalar mesons with symmetric bound state vertex functions on the light front}

\author{George H. S. Yabusaki}
\email{yabusaki@gmail.com} 
\affiliation{Laborat\'orio de F\'\i sica Te\'orica e Computacional, 
Universidade Cruzeiro do Sul, 01506-000, S\~ao Paulo, SP, Brasil}

\author{Ishtiaq Ahmed}
\email{ishtiaq@ncp.edu.pk}
\affiliation{Laborat\'orio de F\'\i sica Te\'orica e Computacional, 
Universidade Cruzeiro do Sul, 01506-000, S\~ao Paulo, SP, Brasil}
\affiliation{National Centre for Physics, Quaid-i-Azam University Campus, 
Islamabad, 45320 Pakistan}

\author{M.~Ali Paracha}
\email{aliphoton9@gmail.com}
\affiliation{Department of Physics, School of Natural Sciences, 
National University of Science and Technology, Islamabad, Pakistan}

\author{J.~P.~B.~C.~de~Melo}
\email{joao.mello@cruzeirodosul.edu.br}
\affiliation{Laborat\'orio de F\'\i sica Te\'orica e Computacional, 
Universidade Cruzeiro do Sul, 01506-000, S\~ao Paulo, SP, Brasil}
 
\author{Bruno El-Bennich}
\email{bruno.bennich@cruzeirodosul.edu.br}
\affiliation{Laborat\'orio de F\'\i sica Te\'orica e Computacional, 
Universidade Cruzeiro do Sul, 
01506-000, S\~ao Paulo, SP, Brasil}
\affiliation{Instituto de F\'isica Te\'orica, Universidade Estadual Paulista, 
01140-070 S\~ao Paulo, SP, Brazil}

\date{\today}

\begin{abstract}

We study the electromagnetic form factors, decay constants and charge radii
of the pion and kaon within the framework of light-front field theory formalism where we use an 
ansatz for the quark-meson interaction 
bound-state function which is symmetric under exchange of quark and antiquark momentum.
The above mentioned observables are evaluated for the $+$ component of the electromagnetic current, $J^+$, in the Breit frame. 
We also check the invariance of these observables in other frames, whereby both the valance and the non-valence contributions 
have to be taken into account, and study the sensitivity of the electromagnetic form factors  and 
charge radius to the model's parameters; namely, the quark masses, $m_u=m_d$, $m_{\bar s}$, and the regulator mass, $m_R$. 
It is found that after a fine tuning of the regulator mass, i.e. $m_R=0.6$ GeV, the model 
is suitable to fit the available experimental data 
within the theoretical uncertainties of both the pion and kaon.
\end{abstract}
\maketitle

\section{Introduction}
\label{intro}

The theory of strong interactions, Quantum Chromodynamics (QCD), has been the object of theoretical and experimental 
scrutiny for four decades now and is perturbatively well defined. Indeed, at large momentum transfer, perturbative
calculations successfully describe a wealth of subatomic phenomena. However, QCD is also a theory whose 
elementary excitations are confined and even in the far infrared below the typical hadronic scale, $E \approx 1$~GeV,
QCD does not seem to break down~\cite{Bashir:2012fs,Cloet:2013jya,Roberts:2015dea}. Thus, nonperturbatively 
QCD may well be rigorously defined but a full solution is yet out of reach: there exists no simple Schr\"odinger picture 
of a many-body Hamiltonian as in quantum mechanics~\cite{Brodsky1998}. This is due to the intrinsic nonperturbative 
nature of quark-antiquark pair creation and annihilation in a relativistic quantum field theory, which entails non-conservation
of particle number.  In hadron physics, this situation leads to seek computational approaches beyond perturbative QCD and 
many models or effective theories have been proposed to tackle QCD in the nonperturbative regime with the aim
to describe hadron phenomenology.

In this context, one possibility to develop a nonperturbative covariant framework is the light-front field theory formalism 
proposed by Dirac in 1949~\cite{Dirac}. In this approach, the hadronic bound states are described by wave functions in the 
light-front space-time hyper surface, defined by the coordinate $x^+=x^0+x^3=0$, and due to the stability of the Fock-state 
decomposition these wave functions are covariant under kinematic boosts~\cite{Brodsky1998,Perry90}.

On the other hand, in understanding the dynamical properties of nonperturbative QCD, the light pseudoscalar mesons
and in particular the pion play a crucial role. Remarkably, the latter is a bound state of massive antiquark-quark pairs as
well as the almost massless Goldstone boson associated with chiral symmetry breaking. There have been many studies of 
their  static properties~\cite{Tobias92,Maris:1997hd,Maris:1997tm,Maris:1999nt,Maris:1999bh,Ebert:2005es} and their 
dynamical properties have also been investigated theoretically~\cite{Lepage:1979zb,Lepage:1980fj,Frederico:1994dx,Melo1999,
Bakker2001,Maris:2000sk,Maris:2002mz,Roberts:2010rn,Desplanques2009,Nguyen:2011jy,Simula2002,Melo2002,Suisso2002,
deMelo:2003uk,deMelo:2005cy,deMelo:2006rg,Dong2012,Loewe2007,Choi2007,Braguta2008,ElBennich:2008qa,
ElBennich:2004px,ElBennich:2004jb,ElBennich:2004mt,Brodsky:2007hb,Raha2009,Raha2010,Paula2010,Albrecht2010,
Leitner2011,Choi,Bakker2014,Biernat2014,Cheng2004,Wilson2000,Melo2004,Melo2006a,Melo2006b,Pereira2007,Silva2012,
Bennich2013,Melo2014,Melo2014v2,Dong2006} and experimentally~\cite{Amendolia1986,Data3, Data2,Data2a,Horn,Tadevosyan,
Amendolia:1986wj,Brauel,Horn:2007ug,Huber:2008id,Blok:2008jy}.

Taking advantage of the simple structure of the Fock space and the vacuum 
in light-front quantization, various hadronic properties 
of bound states, such as decay constants and electromagnetic form factors of the pion, kaon and nucleon, have been calculated
\cite{Melo2002,Melo1999,Cheng2004,Wilson2000,Melo2004,Melo2006a,Melo2006b, Pereira2007,Silva2012,Melo2014,Melo2014v2,
Dong2006,Bennich2013} and successfully compared with their experimental values~\cite{Amendolia1986,Data3,Data2,Data2a,Horn,
Tadevosyan,Amendolia:1986wj,Brauel,Horn:2007ug,Huber:2008id,Blok:2008jy}.  Since the light front component, $J^+$, has been 
successfully employed to calculate electromagnetic form  
factors~\cite{Tobias92,deMelo1997,Simula2002,Dziembowski87,Cardarelli96,Jaus99,Hwang2003,Huang2004}, 
the light-front approach also 
offers a theoretical framework to extract from them useful information on the valence and non-valence components of the meson's 
wave function.

In the present simultaneous study of electromagnetic form factors, charge radii 
and decay constants, we adopt the light-front field 
theory formalism of Refs.~\cite{Tobias92,Frederico:1994dx} wherein the Bethe-Salpeter amplitude of the $q\bar q$ bound states 
was modeled for two different momentum constellations, namely a symmetric~\cite{Melo2002} and nonsymmetric vertex model~\cite{Melo1999}. 
Here, the vertex refers to the $\bar qq$ pair coupling to the pseudoscalar meson in an effective Lagrangian. Using a nonsymmetric vertex model, 
E.~O.~Silva {\em et. al.\/}~\cite{Silva2012} calculated the aforementioned pion and kaon observables which are in agreement with 
experimental data. However, a momentum distribution of the meson that is symmetric under the exchange of the quark and antiquark 
momenta is more realistic and such a model for the Bethe-Salpeter amplitude should improve the description of or at least equally well 
reproduce all observables presented in  Ref.~\cite{Silva2012}.  Thus, we here use the  same component, $J^+$,  of the light-front 
electromagnetic current, though with a symmetric momentum description of quark-meson bound-state vertex. 
It is important to recall here that the choice of $J^+$ with the Drell-Yan condition $q^+=0$ guarantees  that pair-term 
contributions (non-valence terms) vanish~\cite{Melo2002,Melo1999}. On the other hand, to preserve rotational symmetry, the pair  
contribution must be included~\cite{RS1,RS2,RS3,RS4,RS5,RS6}. Consequently, we here employ both the valence and non-valence contributions 
considering the case $q^+\neq0$.

This paper is organized as follows: Sec.~\ref{sec2} serves to summarize the general framework where subsequently the different 
physical observables, namely the electromagnetic form factors, decay constant and charge radii are discussed in turn. 
In Sec.~\ref{sec3}, we present our numerical results and analyze the observables' dependence on variation of the model 
parameters. In the last section, we give our conclusions.

\section{The model \label{sec2}}

In this section, we briefly summarize the model and the computational tools of the light-front formalism required to investigate 
the pseudoscalar meson's electromagnetic form factors, charge radii and the decay constants. Our approach is based on similar 
earlier work~\cite{Melo1999,Melo2002}, where the following effective Lagrangian for the $\bar qq$ bound state was employed:
\begin{eqnarray}
  \mathcal{L}_\mathrm{eff} & = &  -ig\, \vec\phi\!\cdot\bar q\gamma^5\vec\tau q \ , 
\end{eqnarray}
where $g=m_{0^-}/f_{0^-}$ is the coupling constant, $m_{0^-}$ and $f_{0^-}$ denote the mass and decay constant of a pseudoscalar 
meson, respectively, and $\vec \phi$ represents the scalar field. We make a symmetric ansatz for the $\bar qq$-meson vertex 
which describes the bound state,
\begin{align}
\Lambda (k,P) =  &  \ \mathcal{C} \big [ (k^2-m_R^2 + i\epsilon)^{-1}   \nonumber \\
                        + &\   ((P-k)^2-m_R^2+i\epsilon)^{-1} \big ] , 
\label{vertex}
\end{align}
where it is clear that the $\Lambda(k,P)$ is symmetric under the exchange of the quark and antiquark momenta, $k$ and $(P-k)$;
$P$ is the total momentum of the meson and $\mathcal{C} $ a normalization constant. In the following, we discuss the analytic light-front 
formulation  of the electromagnetic form factor, charge radius and decay constant.

\subsection{Electromagnetic form factors}

The covariant electromagnetic form factor, $F^\mathrm{em}_{0^-}$ is defined by a matrix element where the electromagnetic 
current, $J_\mu=e_q\bar q\gamma_\mu q$, is sandwiched between the initial and final bound states of the same meson:
\begin{equation}
 P_\mu\,  F^\mathrm{em}_{0^-}(q^2)=  \langle M_{0^-}(p^\prime)|J_\mu | M_{0^-}(p)\rangle,
 \label{ffactor}
\end{equation}
where $M_{0^-} =  \pi^+, K^+$, $P_\mu= (p+p^\prime)_\mu$ and $q^2=(p-p^\prime)^2$ is the square of the momentum transfer.

The electromagnetic form factor in the impulse approximation is obtained from triangle diagrams, each of which contains 
one spectator quark. In this approximation, the covariant  electromagnetic current of a pseudoscalar meson, $J_\mu$, 
that enters Eq.~(\ref{ffactor}) can be written as follows~\cite{BrodskyB,Coster1994}:
\begin{eqnarray}
 J_\mu & = & N\! \int\frac{d^4k}{(2\pi)^4}\, 
 \mathrm{Tr}\bigg[  \frac{1}{\slashed k-m_{\bar q}+i\epsilon} \gamma^5\frac{1}{\slashed k-\slashed p^\prime-m_q+i\epsilon}
 \gamma_\mu \notag \\
  & \times &\frac{1}{\slashed k-\slashed p-m_q+i\epsilon}\gamma^5   \bigg ]   \Lambda(k,p')\Lambda(k,p)  +  \big [q\leftrightarrow \bar q\big],
 \label{current1}
\end{eqnarray}
with the normalization,  
\begin{eqnarray}
   N  & = &  \frac{-2 \imath\, e_q \hat m_{0^-}^2N_c}{f_{0^-}^2}\ ,  \qquad \hat m_{0^-}~:~\frac{m_q+m_{\bar q}}{2}\ ,  \notag 
 \end{eqnarray}
where~$N_c=3$ is the number of colors and $f_{0^-}$ the pseudoscalar weak decay constant.

The light front-form variables are,
\begin{eqnarray}
 k^+ & = &k^0+k^3, \ k^-=k^0-k^3, \ \vec k_\perp\equiv(k^1,k^2) \notag \\
 q^+ & = & \sqrt{-q^2}\sin\alpha, \ q_x=\sqrt{-q^2}\cos\alpha,\ q_y=0\notag \\
 q^2&=&q^+q^--(\vec q_\perp)^2~. 
 \label{lfvariables}
\end{eqnarray}
Here, we use the Drell-Yan condition,~$q^+=0$, in the Breit frame, which implies $\alpha=0$.
However, our results are frame invariant, i.e. invariant for  $\alpha\neq 0$ where both the valence and the non-valence contributions 
become important. After introducing the front-form variables in Eq.~\eqref{lfvariables} and using $\gamma^+=\gamma^0+\gamma^3$ 
to obtain the $J^+$ component of the current in Eq.~(\ref{current1}), the electromagnetic form factor becomes:
\begin{eqnarray}
 F^\mathrm{em}_{X^{0^-}}(q^2) & = & \frac{N}{P^+ }  \int \frac{d^2k_\perp}{(2 \pi)^4}\, dk^+dk^- \, 
  \mathrm{Tr}\, [\mathcal{O}^+]   \notag \\
 & \times& \Gamma(k^+,p^+,p^{\prime+})+[q\leftrightarrow \bar q] \ ,
 \label{formfactor}
\end{eqnarray}
In Eq.~\eqref{formfactor}, the trace in light-front coordinates is,
\begin{eqnarray}
  \tfrac{1}{4}\, \text{Tr}[\mathcal{O}^+]&=&
  \frac{1}{4}k^+q_\perp^2+(k^+-p^+-p^{\prime +})(k_{\perp}^2-k^+k^-)\notag \\
  & - &   k^-p^+p^{\prime+}-(p^{\prime +}  k_\perp\cdot p_\perp +p^+k_\perp\cdot p_\perp^\prime)
  \notag \\
  & - &k^+(2k_\perp^2+m_{\bar q}^2-2m_qm_{\bar q})-(p^++p^{\prime+})m_qm_{\bar q},\notag \\ 
\label{tracelf}
 \end{eqnarray}
and
\begin{eqnarray}
 \Gamma(k^+,p^+,p^{\prime+})&=&\frac{\Lambda(k^+,p^+) \Lambda(k^+,p^{\prime+})} 
   {(k^2 -m^2_{\bar{q}}+i\epsilon)((p-k)^2 -m^2_{q}+i\epsilon)}\notag \\
    & \times & \frac{1}{((p^\prime-k)^2 -m^2_{q} + i\epsilon)} \ ,
 \end{eqnarray}
where $k^2=k^+(k^--k^-_{\text{on}})$ where $k^-_{\text{on}}$ is the on-energy-shell value of the corresponding momentum given by,
\begin{eqnarray}
  k^-_{\text{on}}&=&\frac{k^2_\perp+m_{\bar{q}}^2}{k^+} \ .
\end{eqnarray}
In terms of light-front variables, the bound-state function in Eq.~\eqref{vertex} becomes,
\begin{eqnarray}
   \Lambda(k^+,p^+)  & = &  \frac{\mathcal{C} }{(p^+-k^+) (p^--k^--\frac{(p-k)^2_\perp+m_R^2-i\epsilon}{p^+-k^+})} \notag \\
     & + & \frac{\mathcal{C} }{k^+(k^--\frac{k^2_\perp+m_R^2-i\epsilon}{k^+})} \ .
\label{vertexlf}
\end{eqnarray}
Collecting all ingredients from Eqs.~\eqref{tracelf}--\eqref{vertexlf}, we insert them in  Eq.~\eqref{formfactor}
and after $k^-$ energy integration (see appendix of Ref.~\cite{Melo2002}) with $x=\frac{k^+}{p^+}$ the electromagnetic form 
factor can be rewritten,
\begin{eqnarray}
   F_{0^-}^{\mathrm{em}}  & = &  \frac{\mathcal{N}}{P^+}    \int{\frac{d^2k_\perp d x }{x(1-x)}\Phi^*(x,k_\perp)
   \Phi(x,k_\perp)}\theta(x)\theta(1-x)\notag \\
    & \times &   \left  [ \tfrac{1}{4}k^+q_\perp^2-k^-_{\text{on}}\, p^+p^{\prime+}-(p^{\prime +}
      k_\perp\cdot p_\perp +p^+k_\perp\cdot p_\perp^\prime) \right ] \notag \\
 \label{ffactor2}
 \end{eqnarray}
where $\mathcal{N}=\frac{N\mathcal{C}^2}{(2\pi)^3}$ and
\begin{eqnarray}
\lefteqn{\Phi(x,k_\perp,p^+,\vec{p}_{\perp}) =  \bigg [\frac{1}{(1-x)(m_{0^-}^2-\mathcal{M}^2(m_q^2,m_R^2))}\notag }\\
 & + & \frac{1}{x(m_{0^-}^2-\mathcal{M}^2(m_R^2,m_{\bar q}^2))} \bigg ] \frac{1}{m_{0^-}^2-\mathcal{M}^2(m_q^2,m_{\bar q}^2)},\notag \\
 &+&[q\leftrightarrow \bar q]\label{wavefunction}
\end{eqnarray}
with
\begin{equation}
 \mathcal{M}^2(m_a^2,m_b^2)=
 \frac{k_\perp^2+m_a^2}{x}+\frac{(p-k)_\perp^2+m_b^2}{(1-x)}-p_\perp^2\notag
\end{equation}
Note that the appearance of the second term of Eq.~\eqref{wavefunction} is due to the symmetric character of the meson-quarks vertex, 
which is absent in Refs.~\cite{Melo1999,Silva2012} where the authors consider a nonsymmetric behavior of the vertex function. 
As it was shown~\cite{Melo1999,Melo2002,Bakker2001}, to preserve general covariance  the non-valence contribution
is mandatory. Thus, in Eq.~\eqref{ffactor2} the step functions $\theta(x)$ and $\theta(1-x)$ delimit the integration interval, $0 < k < p^+$, 
of the valence contribution, whereas the interval, $p^+<k^+<p^{\prime +}$, corresponds to the non-valence contributions to electromagnetic 
current~\cite{Melo2002}.

\subsection{Charge radius and decay constant}
\label{const}

The mean-square electric charge radius of a meson is a relevant quantity and correlated with the electromagnetic form factor,
\begin{equation}
 F_{0^-}^{\mathrm{em}}(q^2)\simeq1-\frac{1}{6} \, \langle r_{0^-}^2\rangle q^2~.
\end{equation}
Differentiation with respect to~$q^2$, of the above equation yields the charge radius,
\begin{equation}
 \langle r_{0^-}^2 \rangle  =  \frac{dF_{0^-}^{\mathrm{em}}}{dq^2} \Bigg |_{q^2=0} \ . 
\end{equation}

A relevant observable and also our main constraint on the model's parameters is given by the weak decay constant,
$f_{0^-}$. The decay constant of a~$q\bar q$,~bound state can be found  from the following matrix element of the partially 
conserved axial-vector current:
\begin{equation}
   \langle 0|A^\mu | {0^-} \rangle=  \imath \, f_{0^-}\, p^\mu \ .
\end{equation}
where $A^\mu=\bar q\gamma^\mu\gamma^5q$, is the axialvector current.  The weak decay constant is given by,
\begin{eqnarray}
   f_{0^-}& = &  \frac{\imath N_c}{f_{0^-}} \int \frac{d^4k}{(2\pi)^4} \text{Tr}\bigg[\slashed p\gamma^5\frac{\slashed k}{k^2-m_q^2+i\epsilon}\notag \\
            & + &\gamma^5\frac{(\slashed k-\slashed p)}{(k-p)^2-m_{\bar q}^2+i\epsilon}\bigg]\Lambda(k,p)
\end{eqnarray}
We make use of the $+$ component of the axialvector current~$A^+$  
and after integration over $k^-$, one obtains the decay constant in terms of the valence component of the model:

\begin{eqnarray}
 f_{0^-} & = &  \frac{\sqrt{N_c}}{4\pi^3} 
 \int\frac{d^2k_\perp dx}{x}[4xm_q+4m_q(1-x)]\Phi(x,k_\perp,m_\pi,\vec{0}).\notag \\
\end{eqnarray}

\section{Numerical Results and Discussion~\label{sec3}}

The model introduced in Sec.~\ref{sec2} contains three free parameters, namely, 
the regulator mass,~$m_R$,  and the two constituent 
quark masses, $m_u =m_d$ and $m_{\bar s}$, where the strange quark mass is taken from the 
study in Ref.~\cite{Suisso2002}.
The main focus of this study is to constrain the parameters of a more realistic bound-state ansatz to 
accommodate the available experimental 
data on the pion and kaon elastic form factors, decay constants and charge radii. In addition, 
it is also instructive and important to 
check the explicit dependence of these observables on the model's parameter. 

Regarding these goals, we know from a previous study~\cite{Melo2002} that the value of 
the regulator mass,~$m_R=0.6$ GeV,  
reproduces well all experimental data on the pion observables mentioned above. 
It is worthwhile to check whether~$m_R=0.6$~GeV 
is also consistent with the kaon data, for which we compute the values of the decay constants 
and charge radii and compare them 
with their experimental values.

\begin{table*}
\begin{ruledtabular}
 \begin{tabular}{|p{2cm} | p{3cm} | p{3cm} |  p{3cm}   p{3cm} }
 \multicolumn{1}{|c|}{Decay Constant} &\multicolumn{1}{|c}{} & \multicolumn{1}{c}{$m_R=0.6$ GeV; $m_{\bar s}=0.44$ GeV} &
\multicolumn{1}{c}{}& \multicolumn{1}{c|}{}\\
\cline{2-5}\multicolumn{1}{|c|}{and} & \multicolumn{1}{|c}{ Pion }  
& \multicolumn{1}{c|}{ }   &
\multicolumn{1}{|c}{Kaon } & 
\multicolumn{1}{c|}{ }\\
\cline{2-5} \multicolumn{1}{|c|}{Charge radius} & \multicolumn{1}{|c|}{$m_u=md=0.22$ GeV} & 
\multicolumn{1}{|c|}{$m_u=md=0.25$ GeV} &
\multicolumn{1}{|c|}{$m_u=md=0.22$ GeV}& \multicolumn{1}{|c|}{$m_u=md=0.25$ GeV}\\
 \cline{1-5}\hline \multicolumn{1}{|c|}{$\mathlarger{\mathlarger{\mathlarger{f_{0^-}}}}$} & \multicolumn{1}{|c|}{$\qquad$93.12 MeV} & \multicolumn{1}{|c|}{$\qquad$101.85 MeV} & \multicolumn{1}{|c|}{$\qquad$110.81 MeV} & \multicolumn{1}{|c|}{$\qquad$113.74 MeV} \\
  \multicolumn{1}{|c|}{$\mathlarger{\mathlarger{\mathlarger{\langle r_{0^-}\rangle}}}$} & \multicolumn{1}{|c|}{$\qquad$0.736 fm} & \multicolumn{1}{|c|}{$\qquad$0.670 fm} & \multicolumn{1}{|c|}{$\qquad$0.754 fm} & \multicolumn{1}{|c|}{$\qquad$0.687 fm} \\
\end{tabular}
\caption{Calculated decay constants and charge radius for two 
light-quark masses and corresponding experimental  values.
The experimental date with errors bar, are; 
$f_\pi^{\text{exp}}=92.42\pm0.021$~MeV,
$\langle r_\pi\rangle^{\text{exp}}=0.672 \pm 0.008$  fm, 
$f_K^{\text{exp}}=110.38\pm 0.1413$ and 
$\langle r_K\rangle^{\text{exp}}=0.560 \pm 0.031$ fm.
Experimental date from~\cite{PDG2014,Amendolia:1986wj,Amendolia1986}.}
\label{tab1}
\end{ruledtabular}
\end{table*}

The calculated values of the  observables listed in Table~\ref{tab1} show that $m_R=0.6$ GeV 
is also a suitable value for the 
kaon. Moreover, for $m_{\bar s}=0.44$~GeV,  $m_u=m_d=0.22$ GeV and $m_R=0.6$ GeV, the deviation 
from the experimental values of $f_{\pi^+}(f_{K^+})$ and  $\langle r_{\pi^+}\rangle(\langle r_{K^+}\rangle)$
are $0.76\%(0.37\%)$ and $8.7\%(34.10\%)$, while with $m_u=m_d=0.25$ GeV, the mismatch is 
$9.3\%(3.02\%)$ and $0.35\%(22.68\%)$, respectively.

\begin{figure}[b]
\vspace*{1.5mm}
\includegraphics[scale=.35,angle=270]{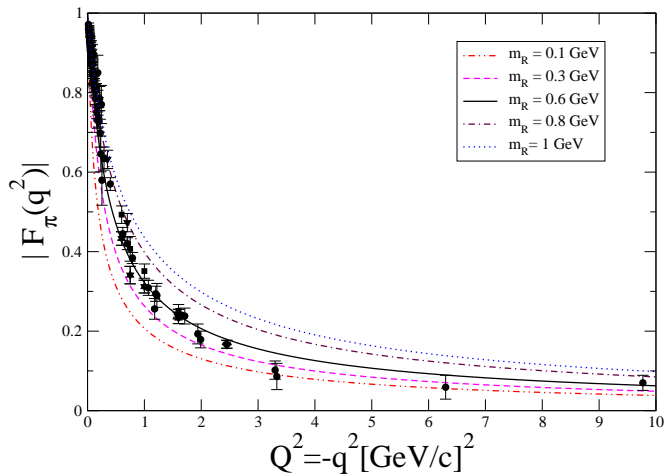} 
\caption{The electromagnetic form factor of the pion as a function of space-like $q^2$.  
The curves correspond to the different values of  
$m_R$ with fixed quark masses: $m_u=m_d=0.220$~GeV. Experimental data: Ref.~\cite{Data2} (circle), 
Ref.~\cite{Data3} (square), Ref.~\cite{Horn} (diamonds), 
Ref.~\cite{Tadevosyan} (up triangle), Ref.~\cite{Huber:2008id} (down triangle).}
\label{fig1} 
\end{figure}

\begin{figure}[b]
\vspace*{1.5mm}
\includegraphics[scale=.35,angle=270]{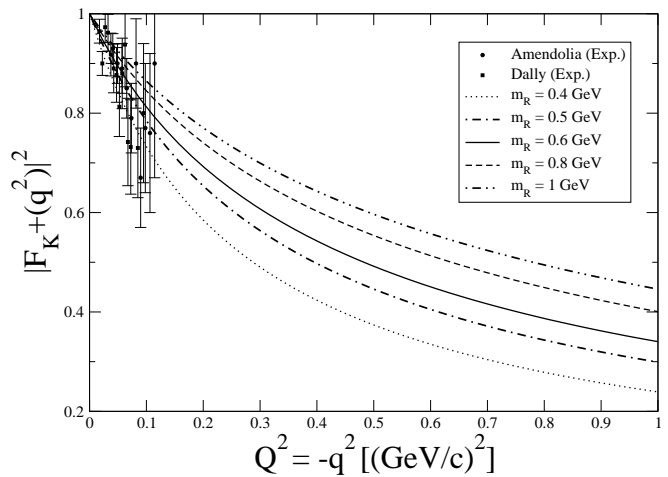}
\caption{The electromagnetic form factor of the kaon as a 
function of space-like $q^2$. The curves correspond to the different values of  
$m_R$ with fixed quark masses: $m_u=m_d=0.220$~GeV, $m_{\bar s}=0.44$~GeV. 
Experimental data: Ref.~\cite{Dally} (square), 
Ref.~\cite{Amendolia1986} (circle).}
\label{fig2} 
\end{figure}

\begin{figure}[t]
\vspace*{-3mm}
\includegraphics[scale=.35,angle=270]{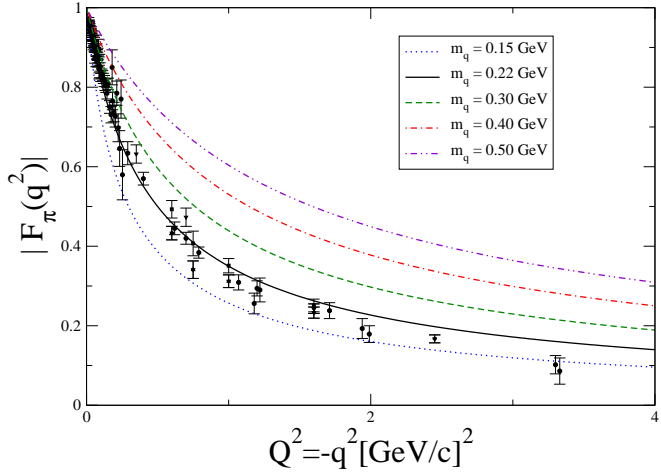}  
\caption{The electromagnetic form factor of the pion as a function of space-like $q^2$. The curves correspond to different values of  
$m_q=m_u$ with regulator mass $m_R=0.60$~GeV.  Experimental data: Ref.~\cite{Data2} (circle), Ref.~\cite{Data3} 
(square), Ref.~\cite{Horn} (diamonds), Ref.~\cite{Tadevosyan} (up triangle),  Ref.~\cite{Huber:2008id} (down triangle).}
\label{fig3}
\end{figure}

\begin{figure}[t!]
\vspace*{1.5mm}
\includegraphics[scale=.35,angle=270]{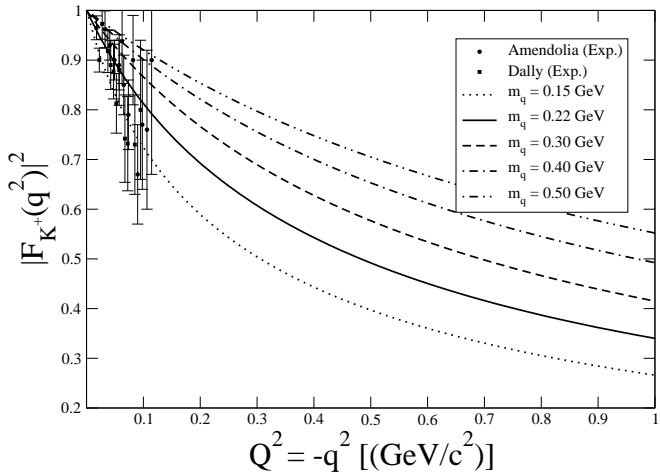} 
\caption{The electromagnetic form factor of the kaon as a function of space-like $q^2$. The curves correspond to 
different values of  $m_q=m_u$ with fixed masses: $m_{\bar s}=0.44$~GeV, $m_R=0.60$~GeV. Experimental data:
solid  circle~\cite{Amendolia1986} (circle) and square~\cite{Dally}.}
\label{fig4}
\end{figure}

\begin{figure}[t]
\vspace*{1.5mm}
\includegraphics[scale=.35,angle=270]{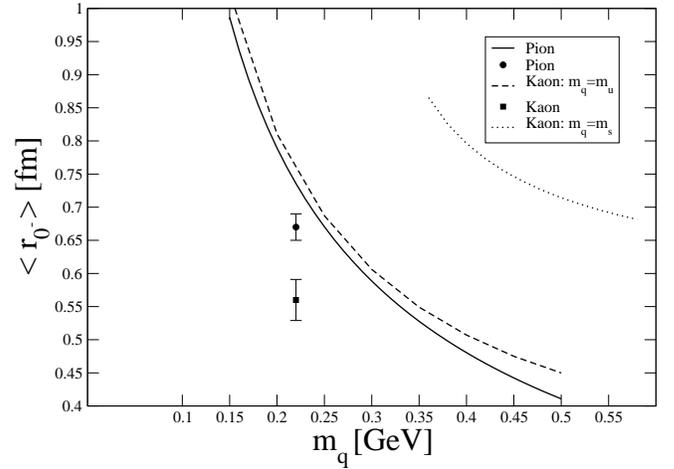}  
\caption{Charge radii $\langle r_{0^-} \rangle$ of the pion and kaon as a function of the constituent 
quark mass $m_u=m_d$ with $m_{\bar s}=0.44$~GeV and fixed regulator mass $m_R=0.6$~GeV. 
The solid circle~\cite{Amendolia:1986wj} and square~\cite{Amendolia1986} are the experimental values 
for the charge radii of the pion and kaon, respectively. }
\label{fig5}
\end{figure}

\begin{figure}[t]
\vspace{1.5mm}
\includegraphics[scale=.35,angle=270]{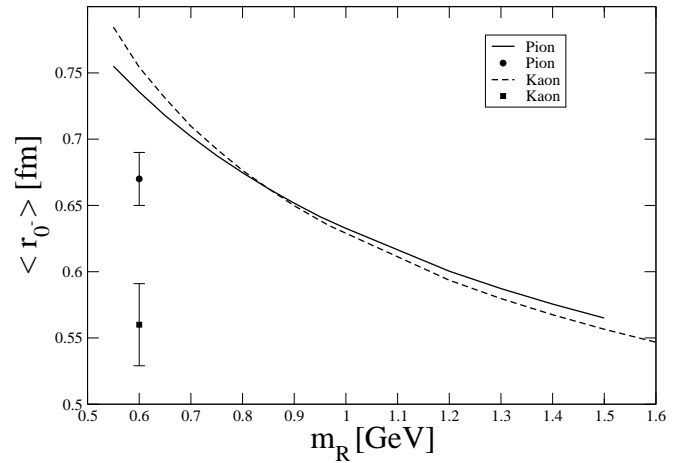} 
\caption{Charge radii $\langle r_{0^-} \rangle$ of the pion and kaon as a function of the regulator mass $m_R$ 
with $m_u=m_d=0.22$~GeV and $m_{\bar s}=0.44$~GeV fixed. The solid circle~\cite{Amendolia:1986wj} and the 
square~\cite{Amendolia1986} are the experimental charge radii values of the pion and kaon, respectively. }
\label{fig6}
\end{figure}

\begin{figure}  
 \vspace*{-3mm}
\includegraphics[scale=.35,angle=270]{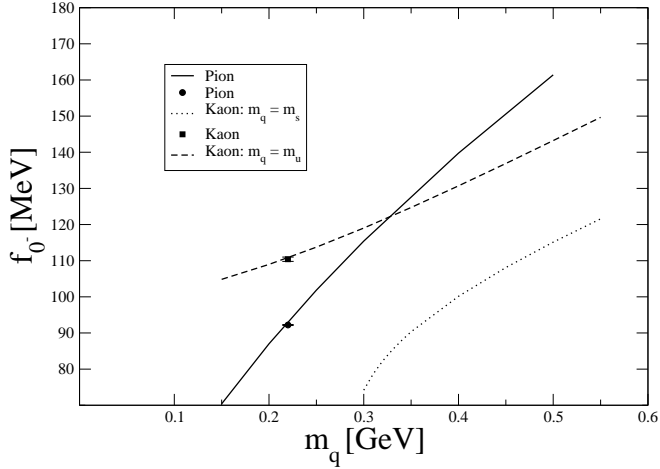}  
\caption{The weak decay constants $f_{0^-}$ of the pion (solid curve with $m_u=m_d=0.22$~GeV) and kaon  (dashed curve: 
$m_{\bar s}=0.44$~GeV) as a function of $m_u=m_d$; dotted curve: kaon decay constant as a function of $m_{\bar s}$.
In all curves $m_R=0.6$~GeV. The solid circle and square are the experimental  decay constants of 
the pion and kaon, respectively~\cite{PDG2014}.}
\label{fig7}
\end{figure}

\begin{figure}[htb] 
\vspace*{-15mm}  
\includegraphics[scale=.35,angle=270]{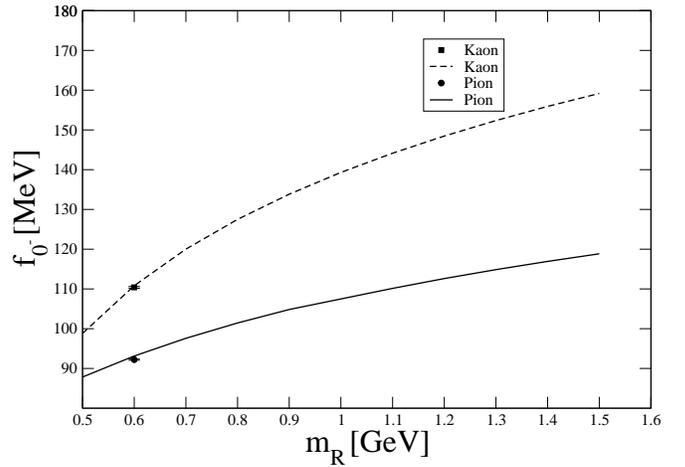} 
\caption{The weak decay constant,  $f_{0^-}$, of the pion (solid curve) and kaon (dashed curve) as a function of the 
regulator mass $m_R$ with $m_u=m_d=0.22$~GeV and $m_{\bar s}=0.44$~GeV fixed. 
The circles and the square are the experimental values or the charge radii of the pion and kaon, respectively~\cite{PDG2014}.}
\label{fig8}
\end{figure}

For comparison, we recall that in case of a non-
symmetric vertex function the ratio of decay constants,
$f_{K}/f_{\pi}$, which measures $SU(3)$ flavor-symmetry breaking, was found to be $1.363$~\cite{Silva2012}. 
Here, we read from the table that $f_{K}/f_{\pi} \simeq 1.189$, which is closer to the measured value 
i.e.,~$f^{\text{exp}}_{K^-}/f_{\pi^-}^{\text{exp}}=1.197\pm0.002\pm0.006$~\cite{PDG2014}.
A graphic representation of the explicit dependence of the form factors,~$F^\mathrm{em}_{0^-}$, decay constants,
$f_{0^-}$ and charge radii~$\langle r_{0^-}\rangle$ on the model's  parameters can be found in Figs.~\ref{fig1} to \ref{fig8}, 
in each of which one parameter is fixed while the other is varied.

In Figs.~\ref{fig1}~and~\ref{fig2}, we plot the electromagnetic form factors of the pion and kaon as a function of $q^2$ for 
various values of $m_R$ and $m_u=m_d$ and $m_{\bar s}$ fixed. As can be seen, the elastic form factors in both figures
are monotonically decreasing functions of $q^2$ with increasing hardness for larger values of $m_R$. 
Fig.~\ref{fig1} also informs us that the available experimental pion data lie in the interval of $0.1$ GeV $\leq m_R\leq1$ GeV and 
from Fig.~\ref{fig2} we deduce that the kaon's experimental form factor data are better reproduced for $m_R \gtrsim 0.5$~ GeV
which coincides with our privileged value, $m_R=0.6$~GeV.

Similarly, in Figs.~\ref{fig3} and \ref{fig4}, the electromagnetic form factors of the pion and kaon are 
plotted as a function of $q^2$ for different values of $m_u=m_d$ whereas $m_R = 0.6$~GeV is fixed.  
One observes a likewise behavior of the electromagnetic form factors, i.e.  $F^\mathrm{em}(q^2)$ becomes harder for increasing 
values of $m_u=m_d$.  We note that for the symmetric vertex function with $m_R=0.6$ GeV, the light-quark mass should be in the 
range $0.15\lesssim m_q\lesssim0.5$~GeV, where the constituent quark mass, $m_q=0.22$~GeV, appears to be the most favorable 
value to accommodate the experimental data.  

Next, the explicit dependence of the charge radii $\langle r_\pi\rangle$ and $\langle  r_K\rangle$ on $m_u=m_d$ is depicted in Fig.~\ref{fig5} from which we deduce  that the charge radius exhibits a nonlinear behavior and decreases strongly for large values 
of $m_u=m_d$, as expected.  For instance, for $m_u=m_d=0.15$~GeV, the size of the pion (kaon) charge radius is about $0.98$~fm
($\sim 1$~fm), whereas for $m_u=m_d=0.5$~GeV this size reduces to $0.43$~fm ($0.45$~fm).  A similar behavior can be seen in
Fig.~\ref{fig6},  where the charge radii are plotted against the regulator mass $m_R$ with $m_u=m_d=0.22$ GeV and 
$m_{\bar s}=0.44$~GeV. However, these effects are less pronounced for variations of $m_R$ than of $m_u=m_d$.

Finally, the weak decay constants, $f_\pi$ and $f_K$, are plotted in Fig.~\ref{fig7} and Fig.~\ref{fig8} as functions 
of $m_u=m_d$ ($m_R=0.6$ GeV) and $m_R$ ($m_u=m_d=0.22$ GeV), where in both cases
 $m_{\bar s} = 0.44$~GeV.
We observe that in contrast to the charge radii  the decay constants  are continuous increasing functions of  $m_q$ and $m_R$.  
Moreover, the charge radii and the decay constants  are  more sensitive to the quark mass values, $m_q$, than to $m_R$.
It is worthwhile to point out, as discussed in Ref.~\cite{Silva2012} for the non-symmetric vertex, that the sensitivity 
of the kaon decay constant to the strange quark mass is very modest, whereas  for the present symmetric vertex the 
$m_{\bar s}$ dependence is quite significant.

We stress that the regulator mass $m_R=0.6$ yields the best fit to the experimental values of the decay constants and 
charge radii as discussed above. Furthermore, the decreasing (increasing) behavior of  the charge radii (decay constants) 
with $m_q$, as depicted in Figs.~\ref{fig5} and \ref{fig7}, satisfies Tarrach's relation~\cite{Tarrach:1979ta},
 i.e.  $\langle \, r_{0^-}\rangle\sim1/m_q$ and $f_{0^-}\sim1/ \langle \, r_{0^-}\rangle$. 

\newpage 

\section{Conclusions}\label{conc}

We reexamined the light-front approach to the light pseudoscalar 
mesons~\cite{Melo1999,Melo2002,Silva2012} by 
considering a symmetric $q\bar q$ bound-state function~\cite{Melo2002}. 
In this framework and with the given symmetric 
ansatz for the bound-state vertex function, we calculated the charge radii, 
$\langle \, r_\pi \rangle$, and $\langle \, r_K \rangle$, 
weak decay  constants, $f_\pi$ and $f_K$ and the electic form factors, 
$F^\mathrm{em}(q^2)_\pi$ and $F^\mathrm{em}(q^2)_K $.

To constrain our model parameters, namely, the quark masses $m_u$, $m_d$ and $m_s$ 
and the regulator mass $m_R$, we first 
adjusted  their values to repoduce the experimental weak decay constants. 
In doing so, we imposed the same regulator mass for both 
the pion and kaon and found that $m_R=0.6$ GeV is a suitable value 
to describe all experimental data on 
$F^{em}_{\pi(K)}$, $\langle r_{\pi(K)}\rangle$ and $f_{\pi(K)}$ within the 
reasonable theoretical uncertainties.
The numerical results also show this model significantly breaks down for $m_R\geq 1$~GeV, which was already demonstrated 
in the the case of a nonsymmetric vertex function~\cite{Silva2012}.

Moreover, the explicit dependence of charge radii and  decay constants on the quark masses, $m_u=m_d$ and $m_{\bar s}$, 
not only satisfies Tarrach's relation but also favors the range of mass values commonly chosen within the light-front model. 
In addition, by using the privileged values of the model's parameters $m_u=m_d=0.22$ GeV, $m_{\bar s}=0.44$ GeV and 
$m_R=0.6$ GeV,  we find that the pion-to-kaon decay constant ratio is in excellent agreement with its experimental value, 
i.e. the mismatch is barely $0.67\%$. Lastly, the present numerical investigation suggests that these parameter values 
could also be used to study dynamical properties of other pseudoscalar and vector mesons or may apply to heavy-to-light 
transition form factors.

\section*{Acknowledgments}

This work was supported by the Brazilian agencies FAPESP (Funda\c{c}\~ao de Amparo \`a Pesquisa do
Estado de S\~ao Paulo),  CAPES (Coordena\c c\~ao de Aperfei\c coamento de Pessoal de N\'ivel Superior 
and CNPq (Conselho Nacional de Desenvolvimento Cient\'\i fico e Tecnol\'ogico).

\end{document}